# CVC: The Contourlet Video Compression algorithm for real-time applications

Stamos Katsigiannis, Georgios Papaioannou, and Dimitris Maroulis

*S. Katsigiannis and D. Maroulis are with the Real-time Systems and Image Analysis Group, Department of Informatics and Telecommunications, National and Kapodistrian University of Athens, Panepistimioupolis, Ilisia, 15703 Athens, Greece.*

*G. Papaioannou is with the Computer Graphics Group, Department of Informatics, Athens University of Economics and Business, 76 Patission Str., 10434, Athens, Greece.*

Abstract. Nowadays, real-time video communication over the internet through video conferencing applications has become an invaluable tool in everyone's professional and personal life. This trend underlines the need for video coding algorithms that provide acceptable quality on low bitrates and can support various resolutions inside the same stream in order to cope with limitations on computational resources and network bandwidth. In this work, a novel scalable video coding algorithm based on the contourlet transform is presented. The algorithm utilizes both lossy and lossless methods in order to achieve compression. One of its most notable features is that due to the transform utilised, it does not suffer from blocking artifacts that occur with many widely adopted compression algorithms. The proposed algorithm takes advantage of the vast computational capabilities of modern GPUs, in order to achieve real-time performance and provide satisfactory encoding and decoding times at relatively low cost, making it suitable for applications like video conferencing. Experiments show that the proposed algorithm performs satisfactorily in terms of compression ratio and speed, while it outperforms standard methods in terms of perceptual quality on lower bitrates.

# 1 Introduction

Real-time video communications over computer networks are being extensively utilised in order to provide video conferencing and other video streaming applications. Transmitting and receiving video data over the internet and other heterogeneous IP networks is a challenging task due to limitations on bandwidth and computational power, as well as caching inefficiency. The increased cost of computational and network resources underlines the need for highly efficient video coding algorithms that could be successfully utilised under the limited resources of casual users, i.e. commodity devices with limited bandwidth.

The most desirable characteristic of such an algorithm would be the ability to maintain satisfactory visual quality while achieving enough compression in order to meet the low bandwidth requirement. Low computational complexity and real-time performance would also be an advantage since it would allow the algorithm to be used in a wide variety of less powerful devices. Especially in the case of video conferencing, real-time encoding and decoding is necessary. Taking into consideration the diversity of available hardware and bandwidth among users, the algorithm should also have the ability to adapt to the network's end-to-end bandwidth and transmitter/receiver resources.

State of the art video compression algorithms that are commonly utilised for stored video content cannot achieve both optimal compression and real-time performance without the use of dedicated hardware, due to their high computational complexity. Moreover, they depend on multipass statistical and structural analysis of the whole video content in order to achieve optimal quality and compression. This procedure is acceptable for stored video content but cannot be performed in cases of live video stream generation as in the case of video conferencing. Another disadvantage of popular Discrete Cosine Transform (DCT)-based algorithms like H.264 [1], VP8 [2], other algorithms belonging to the MPEG family, etc. is that at low bitrates they introduce visible block artifacts that distract the viewers and degrade the communication experience. In order to cope with this problem, deblocking filters are utilised either as an optional post processing step or as part of the compression algorithm. Nevertheless, in both cases this constitutes a separate processing layer that increases the computational complexity.

Most state of the art video and image encoding algorithms require the transformation of visual information from the spatial (pixel) intensity to the frequency domain before any further manipulation (e.g. JPEG [3], JPEG2000 [4], MPEG1&2 [5][6], H.264 [1], VP8 [2]). Frequency domain methods like the Fourier Transform, the Discrete Cosine Transform (DCT), and the Wavelet Transform (WT) have been extensively exploited in image and video encoding, as they offer advantages such as reversibility, compact representations, efficient coding, capturing of structural characteristics, etc [7].

In this work, we extend the proof-of-concept contourlet-based video compression algorithm that was introduced in [8], and propose CVC, a complete algorithm that is able to achieve increased compression efficiency while minimizing the loss of visual quality. The proposed algorithm is based on the Contourlet Transform (CT) [9], a reversible transform that offers multiscale and directional decomposition, providing anisotropy and directionality [9], features missing from traditional transforms like the DCT and WT. The CT has been utilised in a variety of image analysis applications, including denoising [10][11], medical and natural image classification [12], image fusion [13], image enhancement [14], etc. An important aspect of the CT is its efficient parallel implementation. Despite its computational complexity, by taking advantage of the fast parallel calculations performed on modern graphics processing units (GPUs), a GPU-based encoder is able to provide all the CT's benefits, while achieving real-time performance in commodity hardware.

The proof-of-concept algorithm performed real-time video encoding, targeting content obtained from low resolution sources like web cameras, surveillance cameras, etc. and experiments showed promising results. Although it provided satisfactory visual quality, it lacked some basic elements that would provide sufficient compression efficiency. The most important observation is that the previous version performed the CT on the luminance channel and only subsampled the chrominance channels. This handling of the chrominance channels was insufficient for high frame quality at low bitrates, however. In this paper, the revised compression scheme (CVC) performs the CT decomposition on all channels and combines this with chrominance subsampling. Additionally, the lack of a variable-length entropy encoding scheme did not allow it to compete in terms of compression with current state-of-the-art algorithms, a problem solved in the proposed version by the utilization of the DEFLATE algorithm. Combined with an additional lossless filtering step that enhances the compressibility of the data and a mechanism for producing scalable or non-scalable transmission streams, CVC can achieve increased versatility and efficiency in terms of compression. Moreover, CVC introduces a new quantization strategy for the components of the video stream in order to address the limitations of the previous scheme that only supported three quality levels by utilising custom quantization matrices targeting low, medium and high visual quality. Finally, temporal redundancy is now addressed using motion estimation techniques and CVC's performance is evaluated against other video compression algorithms.

The rest of this paper is organised in four sections. Section 2 provides some important background knowledge, while the proposed algorithm is presented in detail in Section 3. The experimental study for its evaluation is provided in Section 4 and finally conclusions are drawn in Section 5.

## 2 Background

### 2.1 The Contourlet Transform

The Contourlet Transform (CT) [9] is a multiscale image representation scheme that provides directionality and anisotropy, and is effective in representing smooth contours in different directions on an image. The method is realized as a double filter bank consisting of first, the *Laplacian Pyramid* (LP) [15] and subsequently, the *Directional Filter Bank* (DFB) [16]. The DFB is a 2D directional filter bank that can achieve perfect reconstruction and the simplified DFB used for the CT decomposes the image into $2^l$ subbands with wedge-shaped frequency partitioning [17], with $l$ being the level of decomposition. In each LP level, the image is decomposed into a downsampled lowpass version of the original image and a more detailed image, containing the supplementary high frequencies, which is fed into a DFB in order to capture the directional information. The combined result is a double iterated filter bank that decomposes images into directional subbands at multiple scales and is called the *Contourlet Filter Bank* (Fig. 1). This scheme can be iterated continuously in the lowpass image and provides a way to obtain multiscale decomposition. In this work, the Cohen and Daubechies 9-7 filters [18] have been utilised for computing the LP and DFB as proposed in [9].

The energy distribution of the CT coefficients is similar to those of the WT, i.e. most of them are near zero except those located near image edges [19]. Although the CT is considered as an over-complete transform that introduces data redundancy, it provides subbands with better or comparable sparseness with respect to the WT, as shown in Fig. 2, making it suitable for compression.

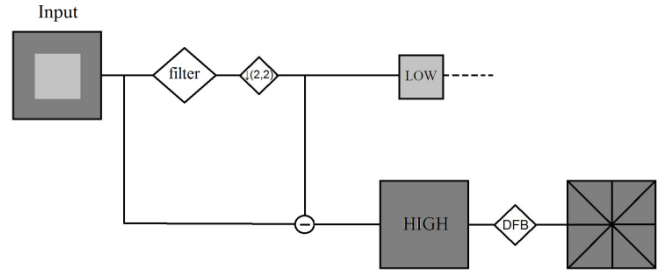

Fig. 1. The Contourlet Filter Bank

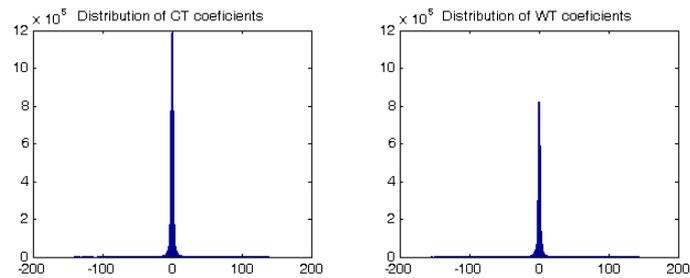

Fig. 2. Distribution of the contourlet and wavelet coefficients of the highpass subbands for the Y channel of the images from the Canon image set.

**2.2 General purpose GPU computing**

Modern personal computers and tablets are commonly equipped with powerful parallel graphics processors (GPUs), whose cores, although individually slower and less sophisticated than the ones of the CPUs, are available in far larger numbers. GPUs are ideal for performing massively parallel computations, especially when inter-thread data exchange can be brought to a minimum. They are commodity hardware and are easy to write optimized code for. The highly parallel structure of GPUs makes them more effective than general-purpose CPUs for algorithms where processing of large blocks of data is done in parallel and in isolation.

In this paper, the most computationally intensive steps of the proposed algorithm are processed on the GPU. For our GPU implementation, the NVIDIA Compute Unified Device Architecture (CUDA) [20] has been selected due to the extensive capabilities and optimized API it provides.

## 2.3 The YCoCg colour space

The RGB-to-YCoCg transform decomposes a colour image into luminance (Y), orange chrominance (Co) and green chrominance (Cg) components. It was developed primarily to address some limitations of the different YCbCr colour spaces and has been shown to exhibit better decorrelation properties than YCbCr and similar transforms [21]. The YCoCg colour space has been utilised in various applications like the H264 compression algorithm, on a real-time RGB frame buffer compression scheme using chrominance subsampling [22], etc. The transform and its reverse are calculated by the equations shown in Table I.

Table 1 RGB to YCoCg transform

| RGB to YCoCg | | YCoCg to RGB | |
|---|---|---|---|
| $Y = \frac{R}{4} + \frac{G}{2} + \frac{B}{4}$ | (1) | $R = Y + Co - Cg$ | (4) |
| $Co = \frac{R}{2} - \frac{B}{2}$ | (2) | $G = Y + Cg$ | (5) |
| $Cg = -\frac{R}{4} + \frac{G}{2} - \frac{B}{4}$ | (3) | $B = Y - Co - Cg$ | (6) |

The Co and Cg components require higher precision than the RGB components in order for the reverse transform to be perfect and avoid rounding errors. Nevertheless, experiments on the Kodak and Canon image sets [23], as well as on 963 outdoor scene images [24] showed that using the same precision for the YCoCg and RGB data when transforming from RGB to YCoCg and back results to an average PSNR of more than 56.87 dB for typical natural images. This quality loss cannot be perceived by the human visual system.

# 3 Algorithm overview

## 3.1 RGB to YCoCg transform and Chroma subsampling

Assuming that input frames are encoded in the RGB colour space with 8 bit colour depth for each chrominance channel, the first step of the algorithm is the conversion from RGB to YCoCg colour space.

The next step is the subsampling of the chrominance channels. It is well established in literature that the human visual system is significantly more sensitive to variations of luminance compared to variations of chrominance. Exploiting this fact, the encoding of the luminance channel of an image with higher accuracy than the chrominance channels provides a low complexity compression scheme that achieves satisfactory visual quality, while offering significant compression. This technique, commonly referred to as chroma subsampling, is utilised by various image and video compression algorithms. In this work, the chrominance channels are subsampled by a user-defined factor $N$ that directly affects the output's visual quality and compression.

Then, the Co and Cg channels are normalized to the range [0, 255]. Since the range of values for the Co and Cg channels is [-127 , 127] for images with 8 bit depth colour per channel in the RGB colour space, as can be derived by the equations 2-3, normalization is performed very fast by adding 127 to every element. For the reconstruction of the chrominance channels at the decoding stage, the Co and Cg channels are normalized to their original range of values by subtracting 127 from each element and the missing chrominance values are replaced by using bilinear interpolation from the four neighboring pixels.

## 3.2 Motion estimation

Motion estimation is performed on the luminance channel of the video in order to exploit the temporal redundancy between consecutive frames. The luminance channel is divided into blocks of size 16x16 and a full search block-matching algorithm is utilised in order to calculate the motion vectors between the current and the previous frame. The maximum displacement between a block and a target block is set as $W$ pixels, leading to $(2W+1)^2$ possible blocks including the original block, with $W$ being user defined. The Mean Squared Error (MSE) is used as the similarity measure for comparing blocks. Although the full search algorithm is an exhaustive search algorithm that significantly increases the computational complexity, it can be efficiently computed in parallel using the capabilities of the GPU and thus is selected due to its simplicity. Contrary to traditional video coding algorithms, the motion-compensated prediction is not applied on the spatial domain (i.e. the raw frame) at this step, but on the contourlet domain, as explained on Section 3.7.

### 3.3 Contourlet Transform decomposition

At this step, all channels are decomposed using the CT. Decomposing into multiple levels using the LP provides the means to separately store different scales of the input that can be in turn individually decoded, thus providing the desirable scalability feature of the algorithm, i.e. multiple resolutions inside the same video stream. At each level, CT decomposition provides a lowpass image of half the resolution of the previous scale and the directional subbands containing the supplementary high frequencies. For example, decomposition of 2 levels of a VGA (640x480) video provides a video stream containing the VGA, QVGA (320x240) and QQVGA (160x120) resolutions. The final result contains the lowpass image of the coarsest scale and the directional subbands of each scale. This property allows CVC to adapt to the network's end-to-end bandwidth and transmitter/receiver resources. The quality for each receiver can be adjusted by just dropping the encoded information referring to higher resolution than requested, without the need to re-encode the video frames at the source. Moreover, due to the downsampling of the lowpass component at each scale, it provides enhanced compression.

### 3.4 Normalization of lowpass component

The lowpass component provided by the CT decomposition is subsequently normalized to the range [0 , 255] in order for its values to fit to 8 bit variables. Experiments on a large number of images from various datasets showed that precision loss due to this normalization is negligible or non-existent depending on the method utilised. It must be noted that during the decoding phase, the lowpass component has to be restored to its correct range.

## 3.5 CT coefficients reduction

In order to compress and transmit/store only perceptually significant information and therefore decrease the bandwidth demands further, the number of different values for the CT coefficients of the directional subbands, as well as for the elements of the lowpass component, is reduced by means of quantization. Quantization is achieved through dividing each element by a user-defined parameter and then rounding to the integer, as explained on the next step of the algorithm. Two separate quantization parameters are used: QPH for the directional subbands and QPL for the lowpass component. QPH takes values in the range [1 , 181]. A QPH of 1 means that no element of the directional subbands is altered, while when QPH=181, all the directional subband elements are reduced to zero. Intermediate values determine the range of the quantized values that can be assigned to CT coefficients. QPL takes values in the range of [1 , 71], with a QPL of 1 meaning that no element of the lowpass component is altered. Although QPL values higher than 71 are possible, the visual quality degradation is severe, as shown in Fig. 4. As a result, the value 71 has been selected as the maximum QPL value. The maximum QPH was selected through experimental evaluation.

This quantization step reduces the number of distinct values an element inside the video stream can take, effectively increasing the compressibility of the data. Higher valued quantization parameters provide higher compression and worse visual quality, while lower valued parameters provide the opposite result. The effect of the quantization procedure in terms of quality is shown in Fig. 3 and 4.

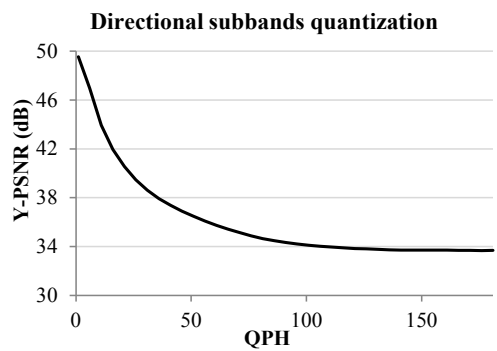

Fig. 3. Effect of QPH on visual quality in terms of Y-PSNR for the "Akiyo" QCIF sequence [29]. Encoder parameters were those used for the creation of the rate-distortion graphs and QPL was set to 1.

## 3.6 Precision change

Experiments with the precision allocated for the contourlet coefficients showed that the CT exhibits resistance to quality loss due to loss of precision in the CT coefficients representation. Exploiting this fact, the precision of the contourlet coefficients is reduced by means of rounding to the nearest integer, resulting in all the components fitting into 8 bit variables. Experiments on the Kodak and Canon image sets showed that rounding to the integer provides an average PSNR equal to 59.6 dB when only the directional subbands' coefficients are rounded and 58.93 dB when also rounding the lowpass content. In both cases, the loss of quality is considered as insignificant due to the fact that it cannot be perceived by the human visual system. As a result, all components are rounded to the integer.

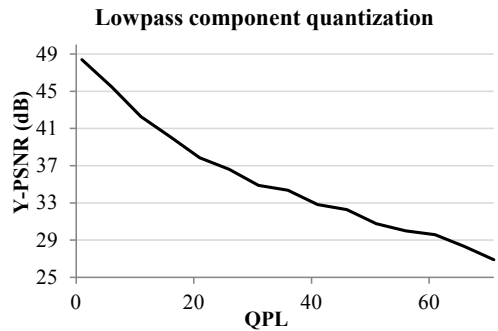

Fig. 4. Effect of QPL on visual quality in terms of Y-PSNR for the "Akiyo" QCIF sequence [29]. Encoder parameters were those used for the creation of the rate-distortion graphs and QPH was set to 1.

## 3.7 Frame types and Migration to CPU

Consecutive frames tend to have small variations and many identical regions, especially in video sequences with static background. This fact can be exploited in order to achieve increased compression efficiency. The algorithm separates video frames into two categories; key frames (K-frames) and predicted frames (P-frames). K-frames are frames that contain all the data for their decoding, while P-frames are the frames between two consecutive K-frames. The number of P-frames between two K-frames is a user defined parameter that affects the total number of K-frames inside a video stream. P-frames are encoded similarly to K-frames up to this step.

For the K-frames, a simple lossless filtering method is applied in order to enhance the compressibility of the lowpass component. At each lowpass component's column, each element's value is calculated by subtracting the value of the element above. Neighboring elements tend to have close values. As a result, a filtered column will have elements with values clustered near zero, providing more compressible data due to the large number of zero-valued sequences of bits inside the bit stream. Filtering is performed across the columns. An example of gain in compression is shown in Fig. 5.

If a frame is identified as a P-frame, all its components are calculated as the difference between the motion-compensated prediction of the components of the current frame and the respective components of the previous frame. Motion-compensated prediction is applied on the CT components of the frame in order to avoid the emergence of artifacts during the CT reconstruction process at the decoding stage. If the prediction is applied on the raw frame, the calculation of the difference provides artificial edges in the image that produce CT coefficients of significant magnitude. Lowering the precision of these coefficients leads to the emergence of artifacts that significantly distort the output image, especially at higher compression levels. Since the CT coefficients are spatially related [25], this problem is avoided by applying the CT on the original frame and then performing the motion-compensated prediction on each CT component by mapping the motion vectors to the corresponding scale and range for each component.

After performing the frame-type-specific operations, all components are passed from the GPU memory to the main memory. Computations up to this point are all performed on the GPU, avoiding unnecessary memory transfers from the main memory to the GPU memory and vice versa. The input frame is passed to the GPU at the beginning of the encoding process and at this step the encoded components are returned to the main memory. Computations from this point on are executed on the CPU, since they are inherently serial (run-length encoding, DEFLATE). Processing those steps on the GPU would provide worse performance than on the host CPU.

**3.8 Run length encoding**

The quantization of the directional subbands provides large sequences of zero-valued CT coefficients, making run length encoding ideal for their compression. Moreover, due to the small variations and the identical regions between consecutive frames, calculating the difference between the motion-compensated prediction of a frame and the previous frame provides components with large sequences of zero-valued elements. Exploiting this fact, the zero-valued coefficients of the directional subbands of the K-frames and of both the directional subbands and the lowpass component of the P-frames are run-length-encoded along the horizontal direction. Non-zero-valued sequences are efficiently compressed at the next step of the algorithm. At the current step, the lowpass component of the K-frames is not expected to have large sequences of zero-valued elements and is thus not run-length-encoded.

**3.9 DEFLATE**

The last stage of CVC consists of the use of the DEFLATE algorithm [26] in order to encode the frame components. The DEFLATE algorithm is a combination of two lossless data compression algorithms. Data are first compressed using a variant of the LZ77 algorithm [27] in order to eliminate duplicate sequences and then with Huffman coding in order to utilize variable-length entropy-based coding. The reverse procedure at the decoding stage is known as the INFLATE algorithm. Another advantageous characteristic of the DEFLATE algorithm is that it provides a standard bit stream and can be implemented in a patent-free manner. This property makes it easier to develop encoders and decoders for the proposed algorithm and port them to other platforms.

CVC supports two modes of encoding with DEFLATE. The first mode supports the separate transmission of each component, thus providing scalability during transmission. The lowpass component and each of the directional subbands at each scale are encoded separately. Despite leading to non-optimal compression, this serves two purposes: The computational time needed for the DEFLATE algorithm is reduced when small amounts of data are encoded and the scalability of the algorithm is not affected, allowing for separate transmission and decoding of each of the frame's scale. For the second mode, called "Non Transmission

Scalable CVC" (CVC-NTS), all the components of a frame are encoded together using DEFLATE. CVC-NTS does not allow the independent transmission of separate scales of the video stream but each scale can still be decoded separately, thus providing scalability at the decoding stage.

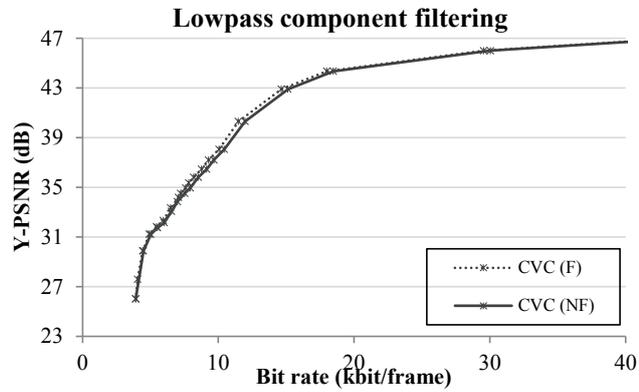

Fig. 5. Rate-distortion graph for the "Akiyo" QCIF sequence when lowpass component filtering is applied (F) and when not applied (NF).

**3.10 GPU implementation**

In order to encode a video frame using CVC, the raw frame's data are first transferred to the GPU memory. Then the next steps of the algorithm are computed on the GPU up to the run-length encoding step, where the encoded components of the frame are transferred to the main memory. Run-length encoding and DEFLATE are inherently serial algorithms and cannot be effectively parallelized. For the GPU implementation of the algorithm, all its steps are mapped to their respective parallel equivalents and can be grouped into four main categories: 1) Operations applied to each element of a 2D array independently, 2) Operations applied to specific chunks of a 2D array, e.g. for every column or row of the array, 3) Operations specific to the Fast Fourier Transform (FFT) calculations needed for computing a 2D convolution, and 4) Operations referring to the motion estimation and motion compensation steps. Resource allocation is done according to the CUDA guidelines in order to exploit the advantages of the CUDA scalable programming model [28].

# 4 Quality and performance analysis

## 4.1 Compression and visual quality

In order to evaluate the performance of CVC, three publicly available video sequences [29], commonly used in video encoding experiments, were encoded. The first two videos (Akiyo, Claire) depict the upper part of the human body and static background in order to resemble video conferencing sequences. The third video (Hall monitor) depicts a hall area with static background and occasional movement of people, resembling a video surveillance sequence. The resolution of the video sequences was QCIF and their frame rate 15 fps.

Rate-distortion graphs for each video sequence reporting the average Y-PSNR and average bit rate per frame are shown in Fig. 6-8. The chrominance channels of the video samples were downsampled by a factor of *N*=4 before being decomposed using the CT and the video stream contained the original, as well as half that resolution. The *W* parameter for motion estimation was set to 8 pixels and the contourlet coefficients of both the luminance and chrominance channels were quantized using the whole range of QPH supported by CVC. For each encoding, QPL was set to $\lfloor QPH/14 \rfloor$ in order to balance lowpass and highpass content degradation. The interval between the K-frames was set to 10 frames. Videos were encoded using both the scalable CVC and CVC-NTS schemes.

The video sequences were also encoded using widely adopted video compression algorithms in order to evaluate the performance of the CVC. The compression algorithms tested were the H261 [30], H263+ [31], and H264 AVC [1]. The libraries included into the FFmpeg framework [32] were utilised for the encoding with the first two algorithms and the x264 encoder [33] was used for H264 AVC. In order to simulate the application scenarios targeted by CVC, i.e. live video creation like video conferencing, all the encoders were set to one-pass encoding, with the same interval between key frames (intra frames) as selected for the CVC, i.e. 10. Additionally, the baseline profile was selected for H264 AVC, without using CABAC [34] due to the real-time restriction. In order to provide comparable results, quantization-based encoding was selected and multiple quantization parameters were tested: the lowest, the highest and intermediate samples. Rate-distortion graphs for each video sample reporting the average Y-PSNR and average bit rate per frame are shown in Fig. 6-8, while sample encoded frames for various bit rates are shown in Fig. 9-12.

As shown in Fig. 6-8, for bit rates up to 6.5 kbit/frame the two CVC schemes achieve comparable or better performance than the other algorithms in terms of visual quality and compression, providing a Y-PSNR between 26 and 32 dB. For bit rates between 6.5 and 10 kbit/frame, their performance is comparable or slightly worse than the other algorithms, providing a Y-PSNR between 32 and 40 dB. For larger bit rates, the proposed algorithm performs better, achieving higher Y-PSNR values at lower bit rates compared to the other algorithms. For bit rates higher than 13.5 kbit/frame the two proposed schemes achieve considerably higher visual quality than the other algorithms. Moreover, as clearly shown in Fig. 12, at low bit rates most of the other algorithms suffer from visible blocking artifacts that significantly degrade the experience of the user. CVC provides more eye-friendly video frames by inherently introducing fuzziness and blurring without the need for an additional deblocking filtering step. Additionally, the CVC-encoded sequences contain more than one resolution that can be separately decoded and transmitted. The actual bitrate utilised depends on which scales the user decides to receive. Containing more than one resolution that can be decoded separately from one another introduces data redundancy and leads to non-optimal compression. Nevertheless, it is a much desired characteristic for a modern video encoding algorithm since it allows the algorithm to adapt to transmitter/receivers resources. Furthermore, the complexity of the proposed algorithm is significantly lower than the H264 AVC algorithm that provided the best results.

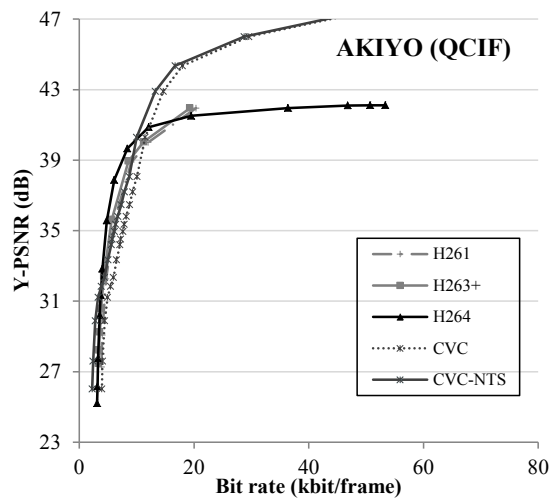

Fig. 6. Rate-distortion graph for the "Akiyo" QCIF sequence, depicting the performance of CVC compared to other algorithms.

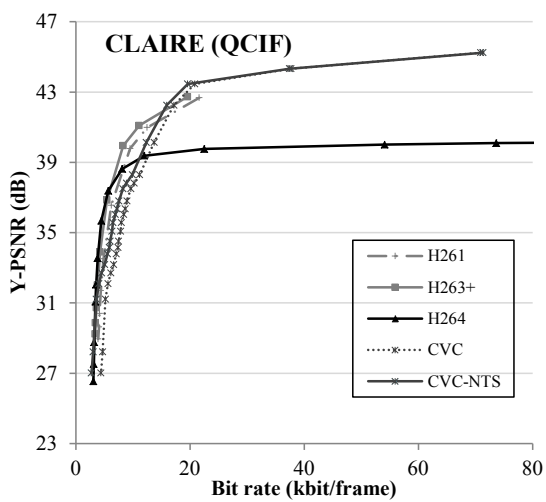

Fig. 7. Rate-distortion graph for the "Claire" QCIF sequence, depicting the performance of CVC compared to other algorithms.

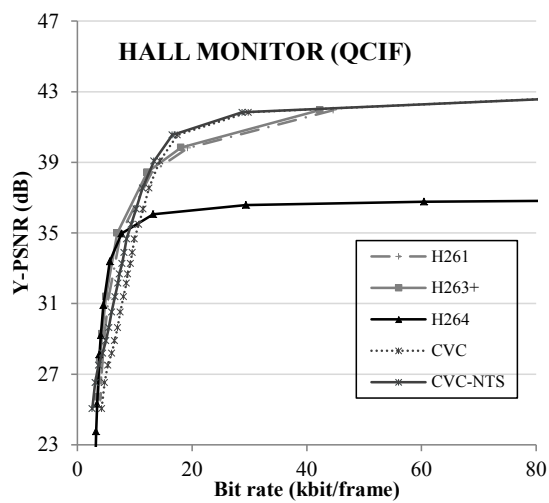

Fig. 8. Rate-distortion graph for the "Hall Monitor" QCIF sequence, depicting the performance of CVC compared to other algorithms.

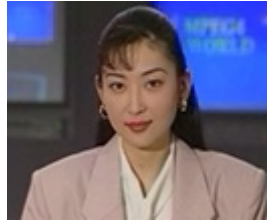 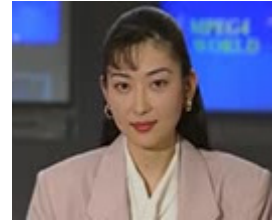

a　　　　　　　　　　　　　　　b

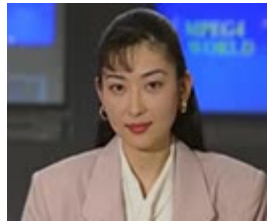 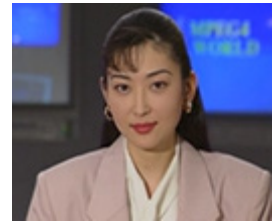

c　　　　　　　　　　　　　　　d

Fig. 9. Cropped sample encoded frames at approximately 20 kbit/frame. (a) CVC-NTS, (b) H261, (c) H263+, (d) H264.

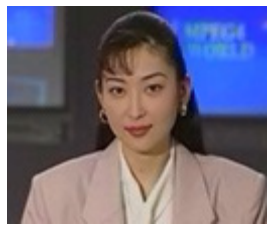 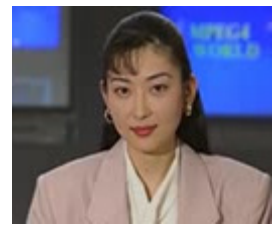

a　　　　　　　　　　　　　　　b

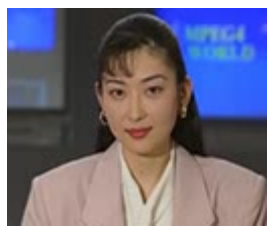 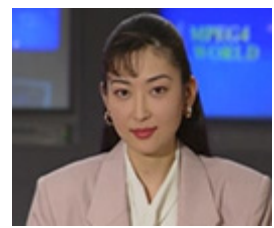

c　　　　　　　　　　　　　　　d

Fig. 10. Cropped sample encoded frames at approximately 11 kbit/frame. (a) CVC-NTS, (b) H261, (c) H263+, (d) H264.

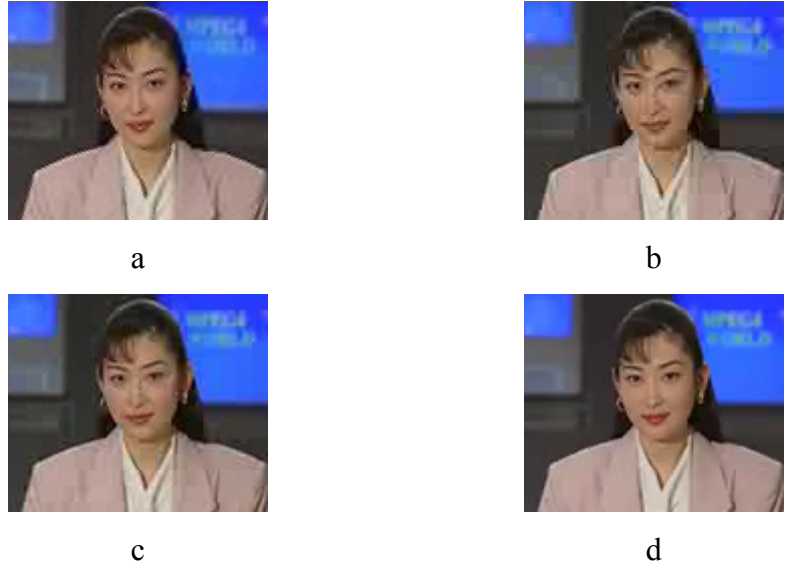

Fig. 11. Cropped sample encoded frames at approximately 4.5 kbit/frame. (a) CVC-NTS, (b) H261, (c) H263+, (d) H264.

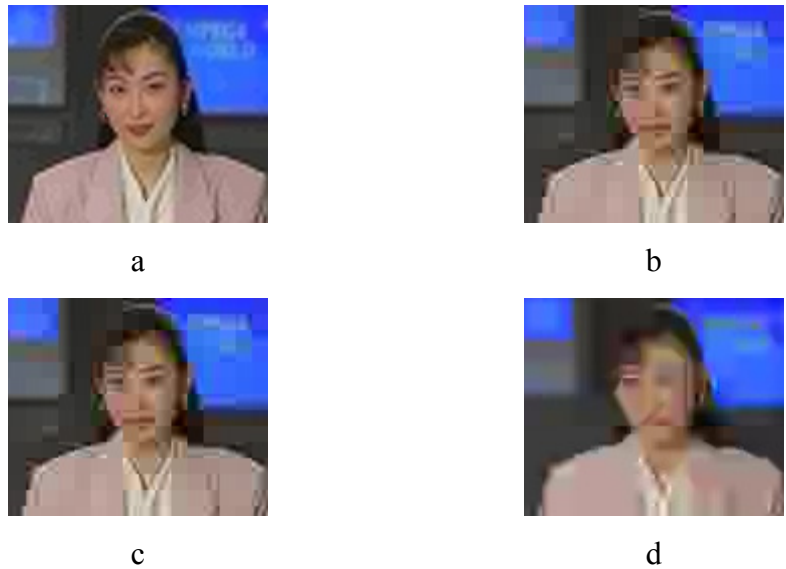

Fig. 12. Cropped sample encoded frames at approximately 3.1 kbit/frame. (a) CVC-NTS, (b) H261, (c) H263+, (d) H264. It must be noted that the H261 and H263+ encoded sequences have average bit rate of 3.9 and 3.3 kbit/frame respectively, since it was the lowest they could achieve under the experimental setting.

### 4.2 Execution times

Fig. 13 presents the average computation time for the basic operations of the encoding and decoding algorithm for a VGA frame for single-threaded execution. The encoded frames contained their original, as well as half that resolution. $N$ was set to 4 and $W$ to 8. Performance tests were conducted on a computer equipped with an Intel Core i3 CPU, 4 GB of DDR3 memory and a NVIDIA GeForce GTX 570 graphics card with 1280 MB of GDDR5 memory. The current implementation can achieve real-time performance for resolutions up to VGA. The use of more powerful or newer graphics cards is expected to provide even faster performance.

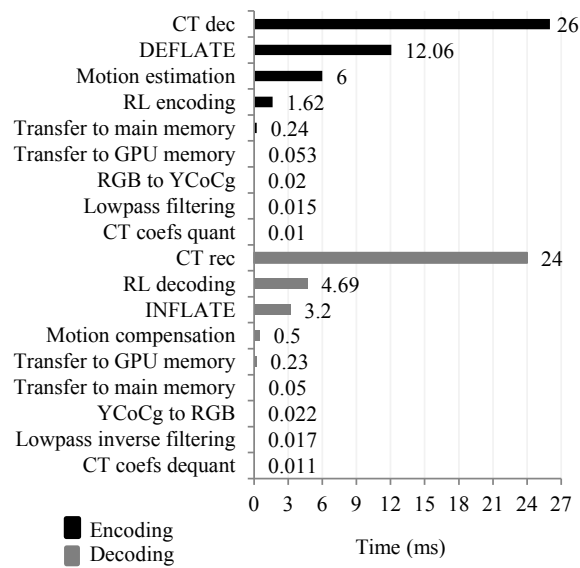

Fig. 13. Average computation times for the basic operations of the encoding and decoding algorithm for a VGA frame.

## 5 Conclusions

In this work, the authors presented CVC, a novel low complexity algorithm for real-time video encoding based on the contourlet transform and optimized for real-time applications. CVC achieves enhanced computational efficiency by migrating its most computationally intensive parts on the GPU. CVC aims on offering a low-complexity alternative for scalable video encoding in the case of real-time applications that can achieve increased visual quality and satisfactory compression, while harnessing the underutilized computational power of modern personal computers.

The incorporation of the DEFLATE algorithm, the CT decomposition of the chrominance channels, the motion estimation, the quantization of CT coefficients and the filtering of the lowpass component provided increased compression efficiency and enhanced visual quality compared to the preliminary version of the algorithm [8]. Furthermore, the new quantization scheme offers support for a wide range of compression and visual quality levels, providing the requisite flexibility for a modern video compression algorithm. The scalable video compression scheme, inherently achieved through the CT decomposition, is ideal for video conferencing content and can support scalability for decoding and transmission, depending on the scheme selected.

CVC achieves comparable performance with the algorithms examined in the case of lower visual quality and higher compression, while achieving comparable or slightly worse performance at medium levels of compression and quality. Nevertheless, it outperforms the examined algorithms in the case of higher visual quality. Moreover, in cases of higher compression, the algorithm does not suffer from blocking artifacts and the visual quality degradation is much more eye-friendly than with other well-established video compression methods, as it introduces fuzziness and blurring, providing smoother images.